\documentclass{desyproc}

\begin{document}
\title{Laboratory Tests of Chameleon Models}

\author{{\slshape Philippe Brax$^1$, Carsten van de Bruck$^2$, Anne-Christine Davis$^3$ and Douglas Shaw$^4$}\\[1ex]
$^1$ Institut de Physique Th\'eorique, CEA, IPhT, CNRS, URA 2306,
  F-91191Gif/Yvette Cedex, France\\
$^2$ Department of Applied Mathematics,
  University of Sheffield Hounsfield Road, Sheffield S3 7RH, United
  Kingdom\\
$^3$ Department of Applied Mathematics and
  Theoretical Physics, Centre for Mathematical Sciences, Cambridge CB3
  0WA, United Kingdom \\
$^4$ Queen Mary University of London, Astronomy Unit,
Mile End Road, London E1 4NS, United Kingdom
}

\acronym{Patras 2009} 

\maketitle

\begin{abstract}
We present a cursory overview of  chameleon models of dark energy and their laboratory tests with an  emphasis  on  optical and Casimir experiments. Optical experiments
measuring the ellipticity of an initially polarised laser beam are sensitive to the coupling of chameleons to photons. The next generation of Casimir experiments may be able to unravel the nature of the scalar force mediated by the chameleon between parallel plates.
\end{abstract}

\section{Chameleon Models}
The observation of the acceleration of the universe has received no definite explanation yet. In fact, there is no convincing explanation for the smallness of the cosmological constant first introduced by Einstein to justify the existence of a spherical and static universe. It would be very natural to assume that there is an underlying mechanism, maybe a symmetry principle, which requires the cancelation of the cosmological constant. If this were the case, then the  existence of a new matter component in the universe could be the explanation to its acceleration. The simplest form for this new type of dark energy is certainly a scalar field. Scalar fields are well-known candidates for the early acceleration phase of the universe. It might be that they also generate the late time acceleration. Typically, runaway models with an inverse power law
\begin{equation}
V= \frac{\Lambda^{4+n}}{\phi^n}
\end{equation}
are interesting candidates as they lead to the acceleration of the universe when the field leaves an attractor solution as $\phi \approx m_{\rm Pl}$. The existence of the attractor guarantees the independence of the late time acceleration from initial conditions. Once the scale $\Lambda$ is appropriately tuned, these models also address the coincidence problem between matter, radiation and dark energy. Unfortunately if the scalar field couples to ordinary matter, its very small mass $m_\phi \approx H_0$ of the order of the Hubble rate now implies that a new fifth force should have been detected. Strong bounds on the coupling
$$\sqrt{8\pi G_N} \alpha= \frac{{\rm d} \ln m_{\rm atom}}{{\rm d} \phi}$$
given by the Cassini experiments $\alpha^2\le 10^{-5}$ prevent the existence of such a coupling. In the absence of a mechanism which allows to decouple dark energy from matter, it seems that dark energy models would be ruled out.

In fact this is not the case thanks to the chameleon mechanism\cite{justin} which allows the scalar field to couple strongly to matter in a sparse environment while being
almost decoupled in a dense setting. Let us consider a scalar-tensor theory with a Lagrangian
$$S= \int d^4 x \sqrt{-g}
(\frac{1}{16\pi G_N}R-\frac{1}{2}(\partial\phi)^2 -V(\phi) +
{\cal L}_m (\psi_m, A^2(\phi) g_{\mu\nu}))$$
The coupling function $A(\phi)$ is responsible for the coupling of the scalar field to matter. Choosing as an example
$$A(\phi)=\exp  \frac{\phi}{M}$$
the coupling to matter  given by
$$\alpha= \frac{m_{\rm Pl}}{M}$$
can be large if $M\le m_{\rm Pl}$. Fortunately, the effective potential felt by the scalar field is not the bare runaway one introduced in the Lagrangian but
$$V_{eff}(\phi)=V(\phi) +\rho_m A(\phi)$$
which depends on the energy density of matter surrounding the scalar field. When this energy density is large, the scalar field is trapped at an effective minimum of the potential where the mass $m_\phi$ is environment dependent and can be very large in very dense regions. This is for instance the case in the atmosphere where the original Galileo experiment was carried out. In this case, the range $\lambda_\phi=m^{-1}_\phi$ of the fifth force mediated by the scalar field is smaller than the shortest detection range of gravitational interactions, i.e. much less than one millimeter. Unfortunately this mechanism is not efficient enough to hide scalar fields away in sparse environments such as the solar system where a strong interaction would lead to large deviations in the planetary motions. This is not the case thanks to a subtle and non-linear effect coined the thin shell effect. When Newton's potential for a  body such as the sun is large enough, the scalar field is effectively trapped inside the body. The absence of any radiated scalar field outside the body implies that the generated scalar force is highly suppressed, hence no deviation in the motion of planets. Dark energy models where this property is present have been called chameleon models\cite{brax1}.

\section{Laboratory Tests}

Like all models invoked to solve cosmological problems, a precise understanding of the nature of chameleon particles would only be achieved by direct detection
in laboratory experiments. The elusiveness of dark energy particles in gravitational experiments has led us to investigate their properties in precision experiments where tiny deviations from the standard model could be detected. Two different types of settings can be envisaged: Casimir and optical experiments\cite{brax2,brax3}. Let us consider the former first. The Casimir effect is one of the triumphs of quantum field theory inasmuch as true quantum fluctuations have  observable macroscopic consequences. Indeed consider two parallel plates of conducting material facing each other at very short distance. The quantum fluctuations of the electromagnetic field induce a power law force between the plates which is attractive and decays like the fourth power of the distance. Remarkably, this effect has been experimentally observed. Now chameleons would certainly generate an extra force between the plates. The force is also of the inverse power law type and is effective in the range $ m^{-1}_c\le d \le m^{-1}_b$ where $m_c$ is the large chameleon mass in the plates and $m_b$ is the smaller chameleon mass in the vacuum between the plates. At short distances the force is nearly constant while it decays exponentially at large distance. The expression of the force is simply
$$\frac{F_\phi}{A} \sim \Lambda^4 (\Lambda d)^{-\frac{2n}{n+2}}$$
where $A$ is the surface of the plates, hence amounting to a pressure. The scale $\Lambda$ is the dark energy energy scale $\Lambda \sim 10^{-3} \rm {eV}$ which corresponds to a small scale
$$\Lambda^{-1} \sim 82 \mu m$$
The algebraic decay of the chameleon pressure is less steep than the algebraic decay of the Casimir force
$$\frac{F_\phi}{F_{\rm cas}}\sim \frac{240}{\pi^2} (\Lambda d)^{\frac{2(n+4)}{n+2}}$$
implying that a detection of the chameleon can only be obtained for scales larger than $\Lambda^{-1}$. In fact, considering possible deviations from the Casimir force, a chameleon would lead to a difference of a few percent for $d=10\mu m$ and would be around a hundred percent for $d=30\mu m$. This is an exciting possibility as
these distances may be probed in the next generation of Casimir experiments.

Optical experiments would probe a different sector of chameleon theories. So far, we have described chameleon theories as resulting from the effective properties
of scalar-tensor theories conformally coupled to matter. In this particular setting, chameleons do not couple directly to photons. A coupling to photons can be introduced
$${\cal L}_{\rm optics}=\frac{e^{\phi/M}} {g^2} F_{\mu\nu}F^{\mu\nu}$$
which breaks conformal invariance. We have chosen the coupling scale $M$ to be the same as the coupling scale to matter. In principle, these two couplings could be different.
The coupling to photons has an important consequence in the presence of an external magnetic field. The chameleon can oscillate into a photon with a probability depending on the coupling strength. This is the Primakov effect and its inverse. This property can lead to observable effects in cavity experiments. When a polarised laser beam enters a cavity where a magnetic field is present, the laser polarisation orthogonal to the magnetic field oscillates into chameleons and vice versa. The lagging effect of the orthogonal polarisation compared the parallel one due to the oscillation into the massive chameleon traveling at a speed smaller than the speed of light implies that the laser light develops a non-vanishing ellipticity.

The mass of the chameleon in the cavity depends on the residual gas density and the magnetic field
$$\rho= \rho_m + \frac{B^2}{2}$$
implying that the coherence length of the laser beam in the cavity becomes magnetic field dependent
$$z_{\rm coh}= \frac{2\omega}{m^2}$$
where $\omega$ is the frequency of the laser light. The mixing angle between the photons and the chameleons is given by
$$\theta= \frac{B\omega}{Mm^2}$$
At the position $z$ in the cavity, the wave function of the orthogonal polarisation is given by
$$\psi(z)= N(1-\frac{1}{N}\sum_{n=0}^{N-1} a_n(z))\cos(\omega z +\frac{1}{N}\sum_{n=0}^{N-1} \delta_n(z))$$
corresponding to $N$ passes of the photons before they leave the cavity through one of the mirrors. The attenuation and the phase shift are given by
$$a_n(z)= 2\theta^2 \sin^2 \frac{m^2(z+nL)}{4\omega},\ \delta_n(z)= \frac{m^2\theta^2}{2\omega} (z+nL)- \theta^2 \sin \frac{m^2(z+nL)}{2\omega}$$
When the cavity length $L=P z_{\rm coh}$ is commensurate with the coherence length, the attenuation and the phase shift simplify
$$a_T=\theta^2,\ \delta_T= \pi \frac{N}{P} \theta^2$$
An important consequence of this result is that the ellipticity of the laser beam after going through the cavity is much larger than the rotation of the
polarisation
$$\frac{\rm ellipticity}{\rm rotation}=\frac{\pi N}{P}$$
Of course, in real experiments the cavity is larger than the interaction length $d$  with the magnetic field. Moreover the chameleon does not reflect instantaneously off the mirrors. This introduces a phase shift
$\Delta_d=\frac{m^2_\phi d}{\omega}$
for the decoherence due to the non-interacting zone and
$\Delta_r  =  \frac{\pi n}{n+2}$
for the reflection. Taking these effects into account, one can give lower bounds on the coupling scale $M$ depending on the expected sensitivities of future experiments.
For instance, with a sensitivity of $10^{-14}$ radians per pass, one would expect a detection of the ellipticity when the coupling is $M\le 10^8$ GeV.

In conclusion, chameleon fields which are motivated by dark energy and its gravitational properties could be within reach in next generation of Casimir and optical experiments.




%


\end{document}